# Split Quaternions and Spacelike Constant Slope Surfaces in Minkowski 3-Space

Murat Babaarslan and Yusuf Yayli

**Abstract.** A spacelike surface in the Minkowski 3-space is called a constant slope surface if its position vector makes a constant angle with the normal at each point on the surface. These surfaces completely classified in [J. Math. Anal. Appl. **385** (1) (2012) 208-220]. In this study, we give some relations between split quaternions and spacelike constant slope surfaces in Minkowski 3-space. We show that spacelike constant slope surfaces can be reparametrized by using rotation matrices corresponding to unit timelike quaternions with the spacelike vector parts and homothetic motions. Subsequently we give some examples to illustrate our main results.

## 1. Introduction

Quaternions were discovered by Sir William Rowan Hamilton as an extension to the complex number in 1843. The most important property of quaternions is that every unit quaternion represents a rotation and this plays a special role in the study of rotations in three-dimensional spaces. Also quaternions are an efficient way understanding many aspects of physics and kinematics. Many physical laws in classical, relativistic and quantum mechanics can be written nicely using them. Today they are used especially in the area of computer vision, computer graphics, animations, aerospace applications, flight simulators, navigation systems and to solve optimization problems involving the estimation of rigid body transformations.

Ozdemir and Ergin [9] showed that a unit timelike quaternion represents a rotation in Minkowski 3-space. Also they expressed Lorentzian rotation matrix generated with a timelike quaternion.

Tosun et al. [10] investigated some properties of one-parameter homothetic motion in Minkowski 3-space. Furthermore, they expressed the relations between velocity and acceleration vectors of a point in moving space with respect to fixed and moving space.

Kula and Yayli [6] showed that the algebra of split quaternions of $H'$ has a scalar product that allows as to identify it with semi-Euclidean space $\mathbf{E}_2^4$. They demonstrated that a pair $q$ and $p$ of unit split quaternions in $H'$ determines a rotation $R_{qp}: H' \to H'$. Moreover they proved that $R_{qp}$ is a product of rotations in a pair of orthogonal planes in $\mathbf{E}_2^4$.

In last few years, the study of the geometry of surfaces in 3-dimensional spaces, in particular of product type $\mathbf{M}^2 \times \mathbf{R}$ was developed by a large number of mathematicians. For instance, constant angle surfaces:

A constant angle surface in $\mathbf{E}^3$ is a surface whose tangent planes make a constant angle with a fixed vector field of ambient space. These surfaces are the generalizations of the concept of helix. These kinds of surfaces are models to describe some phenomena in physics of interfaces in liquid crystals and of layered fluids. Constant angle surfaces were studied in different ambient spaces, e.g. $\mathbf{S}^2 \times \mathbf{R}$ and $\mathbf{H}^2 \times \mathbf{R}$ [2, 3]. Moreover these surfaces were studied in Minkowski 3-space and classified in [6].

Another paper in this direction is [4], where Fu and Yang studied spacelike constant slope surfaces in Minkowski 3-space and classified these surfaces in the same space. They showed that $M \subset \mathbf{E}_1^3$ is a *spacelike constant slope surface lying in the timelike cone* if and only if it can be parametrized by

$$x(u,v) = u \sinh\theta \left( \cosh\xi_1(u) f(v) + \sinh\xi_1(u) f(v) \times f'(v) \right), \qquad (1)$$

where, $\theta$ is a positive constant angle function, $\xi_1(u) = \coth\theta \ln u$ and $f$ is a unit speed spacelike curve on $\mathbf{H}^2$.

Also, $M \subset \mathbf{E}_1^3$ is a *spacelike constant slope surface lying in the spacelike cone* if and only if it can be parametrized by

$$x(u,v) = u \cosh\theta \left( \cosh\xi_2(u) g(v) + \sinh\xi_2(u) g(v) \times g'(v) \right), \qquad (2)$$

where, $\theta$ is a positive constant angle function, $\xi_2(u) = \tanh\theta \ln u$ and $g$ is a unit speed spacelike curve on $\mathbf{S}_1^2$.

In our previous paper [1], we showed a new method to construct constant slope surfaces with quaternions in Euclidean 3-space. In light of recent studies, the purpose of this paper is to give some relations between split quaternions and spacelike constant slope surfaces lying in the timelike cone in Minkowski 3-space. We will show that spacelike constant slope surfaces lying in the timelike cone can be reparametrized by using rotation matrices corresponding to unit timelike quaternions with the spacelike vector parts and homothetic motions. Subsequently we will give some examples to illustrate our main results.

## 2. Basic Notations, Definitions and Formulas

In this section, we give some basic concepts and formulas on split quaternions and semi-Euclidean spaces.

A split quaternion $p$ is an expression of the form

$$p = p_1 1 + p_2 \mathbf{i} + p_3 \mathbf{j} + p_4 \mathbf{k},$$

where $p_1$, $p_2$, $p_3$ and $p_4$ are real numbers and $\mathbf{i}$, $\mathbf{j}$, $\mathbf{k}$ are split quaternion units which satisfy the non-commutative multiplication rules

$$\mathbf{i}^2 = -1, \mathbf{j}^2 = \mathbf{k}^2 = 1, \mathbf{ij} = -\mathbf{ji} = \mathbf{k}, \mathbf{jk} = -\mathbf{kj} = -\mathbf{i} \text{ and } \mathbf{ki} = -\mathbf{ik} = \mathbf{j}.$$

Let us denote the algebra of split quaternions by $\mathbf{H}'$ and its natural basis by $\{1, \mathbf{i}, \mathbf{j}, \mathbf{k}\}$. An element of $\mathbf{H}'$ is called a split quaternion.

For a split quaternion $p = p_1 1 + p_2 \mathbf{i} + p_3 \mathbf{j} + p_4 \mathbf{k}$, the conjugate $\bar{p}$ of $p$ is defined by

$$\bar{p} = p_1 1 - p_2 \mathbf{i} - p_3 \mathbf{j} - p_4 \mathbf{k}.$$

Scalar and vector parts of a split quaternion $p$ are denoted by $S_p = p_1$ and $V_p = p_2 \mathbf{i} + p_3 \mathbf{j} + p_4 \mathbf{k}$, respectively. The split quaternion product of two quaternions $p = (p_1, p_2, p_3, p_4)$ and $q = (q_1, q_2, q_3, q_4)$ is defined as

$$p \times q = p_1 q_1 + \langle V_p, V_q \rangle + p_1 V_q + q_1 V_p + V_p \wedge V_q, \tag{3}$$

where

$$\langle V_p, V_q \rangle = -p_2 q_2 + p_3 q_3 + p_4 q_4$$

and

$$V_p \wedge V_q = \begin{vmatrix} -\mathbf{i} & \mathbf{j} & \mathbf{k} \\ p_2 & p_3 & p_4 \\ q_2 & q_3 & q_4 \end{vmatrix} = (p_4 q_3 - p_3 q_4)\mathbf{i} + (p_4 q_2 - p_2 q_4)\mathbf{j} + (p_2 q_3 - p_3 q_2)\mathbf{k}.$$

If $S_p = 0$, then $p$ is called as a pure split quaternion.

**Definition 1.** $\mathbf{E}^n$ with the metric tensor

$$\langle u, v \rangle = -\sum_{i=1}^{v} u_i v_i + \sum_{i=v+1}^{n} u_i v_i, \ u, v \in \mathbf{R}^n, \ 0 \leq v \leq n$$

is called semi-Euclidean space and denoted by $\mathbf{E}_v^n$, where $v$ is called the index of the metric. If $v = 0$, semi-Euclidean space $\mathbf{E}_v^n$ is reduced to $\mathbf{E}^n$. For $n \geq 2$, $\mathbf{E}_1^n$ is called Minkowski $n$-space; if $n = 4$, it is the simplest example of a relativistic space time [8].

**Definition 2.** Let $\mathbf{E}_v^n$ be a semi-Euclidean space furnished with a metric tensor $\langle , \rangle$. A vector $w \in \mathbf{E}_v^n$ is called

- spacelike if $\langle w, w \rangle > 0$ or $w = 0$,
- timelike if $\langle w, w \rangle < 0$,

- null if $\langle w, w \rangle = 0$ and $w \neq 0$.

The norm of a vector $w \in \mathbf{E}_v^n$ is $\sqrt{|\langle w, w \rangle|}$. Two vector $w_1$ and $w_2$ in $\mathbf{E}_v^n$ are said to be orthogonal if $\langle w_1, w_2 \rangle = 0$ [8].

We can define pseudo-sphere and pseudo-hyperbolic space as follows:

$$\mathbf{S}_v^{n-1} = \left\{ (v_1, ..., v_n) \in \mathbf{E}_v^n \Big| -\sum_{i=1}^{v} v_i^2 + \sum_{i=v+1}^{n} v_i^2 = 1 \right\},$$

and

$$\mathbf{H}_{v-1}^{n-1} = \left\{ (v_1, ..., v_n) \in \mathbf{E}_v^n \Big| -\sum_{i=1}^{v} v_i^2 + \sum_{i=v+1}^{n} v_i^2 = -1 \right\}.$$

Also, for $v = 1$ and $v_1 > 0$, $\mathbf{H}^{n-1} = \mathbf{H}_0^{n-1}$ is called a hyperbolic space of $\mathbf{E}_1^n$.

We can easily check that $-p\bar{p} = -p_1^2 - p_2^2 + p_3^2 + p_4^2$. Therefore we identify $\mathbf{H}'$ with semi-Euclidean space $\mathbf{E}_2^4$. Let $G = \{p \in \mathbf{H}' | p\bar{p} = 1\}$ be the multiplicative group of timelike unit split quaternions. The Lie algebra $\mathfrak{g}$ of $G$ is the imaginary part of $\mathbf{H}'$, that is

$$\mathfrak{g} = \operatorname{Im} \mathbf{H}' = \{p_2 \mathbf{i} + p_3 \mathbf{j} + p_4 \mathbf{k} | p_2, p_3, p_4 \in \mathbf{R}\}.$$

The Lie bracket of $\mathfrak{g}$ is simply the commutator of split quaternion product. Note that the commutation relation of $\mathfrak{g}$ is given by

$$[\mathbf{i}, \mathbf{j}] = 2\mathbf{k}, \quad [\mathbf{j}, \mathbf{k}] = -2\mathbf{i}, \quad [\mathbf{k}, \mathbf{i}] = 2\mathbf{j}.$$

The Lie algebra $\mathfrak{g}$ is naturally identified with $\mathbf{E}_1^3$. Besides, the Lie group $G$ is identified with $\mathbf{H}_1^3$ of constant curvature $-1$ (see [5]).

Thus we can define timelike, spacelike and lightlike quaternions as follows:

**Definition 3.** We say that a split quaternion $p$ is spacelike, timelike or lightlike, if $I_p < 0$, $I_p > 0$ or $I_p = 0$, respectively, where $I_p = p_1^2 + p_2^2 - p_3^2 - p_4^2$ [9].

**Definition 4.** The norm of $p = (p_1, p_2, p_3, p_4)$ is defined as

$$N_p = \sqrt{|p_1^2 + p_2^2 - p_3^2 - p_4^2|}.$$

If $N_p = 1$, then $p$ is called a unit split quaternion and $p_0 = p/N_p$ is a unit split quaternion for $N_p \neq 0$. Also spacelike and timelike quaternions have multiplicative inverses having the property $p \times p^{-1} = p^{-1} \times p = 1$ and they are constructed by $p^{-1} = \bar{p}/I_p$. Lightlike quaternions have no inverses [9].

The vector part of any spacelike quaternions is spacelike but the vector part of any timelike quaternion can be spacelike or timelike. Polar forms of the split quaternions are as below:

(i) Every spacelike quaternion can be written in the form
$$p = N_p (\sinh\theta + \mathbf{v}\cosh\theta),$$
where $\mathbf{v}$ is a unit spacelike vector in $\mathbf{E}_1^3$.

(ii) Every timelike quaternion with the spacelike vector part can be written in the form
$$p = N_p (\cosh\theta + \mathbf{v}\sinh\theta),$$
where $\mathbf{v}$ is a unit spacelike vector in $\mathbf{E}_1^3$.

(iii) Every timelike quaternion with the timelike vector part can be written in the form
$$p = N_p (\cos\theta + \mathbf{v}\sin\theta),$$
where $\mathbf{v}$ is a unit timelike vector in $\mathbf{E}_1^3$ [6, 9].

Unit timelike quaternions are used to perform rotations in the Minkowski 3-space: If $p = (p_1, p_2, p_3, p_4)$ is a unit timelike quaternion, using the transformation law $(p \times V_q \times p^{-1})_i = \sum_{j=1}^{3} R_{ij}(V_q)_j$, the corresponding rotation matrix can be found as

$$R_p = \begin{bmatrix} p_1^2 + p_2^2 + p_3^2 + p_4^2 & 2p_1p_4 - 2p_2p_3 & -2p_1p_3 - 2p_2p_4 \\ 2p_2p_3 + 2p_4p_1 & p_1^2 - p_2^2 - p_3^2 + p_4^2 & -2p_3p_4 - 2p_2p_1 \\ 2p_2p_4 - 2p_3p_1 & 2p_2p_1 - 2p_3p_4 & p_1^2 - p_2^2 + p_3^2 - p_4^2 \end{bmatrix}, \quad (4)$$

where $q = (S_q, V_q)$. We can see that all rows of this matrix are orthogonal in the Lorentzian mean. Therefore the unit timelike quaternion $p = (p_1, p_2, p_3, p_4)$ is equivalent to $3 \times 3$ orthogonal rotational matrix $R_p$ given by Eq. (4). The matrix represents a rotation in Minkowski 3-space under the condition that $\det(R_p) = 1$. This is possible with unit timelike quaternions. Also causal character of vector part of the unit timelike quaternion $p$ is important. If the vector part of $p$ is timelike or spacelike then the rotation angle is hyperbolical or spherical, respectively [9].

In Minkowski 3-space, one-parameter homothetic motion of a body is generated by the transformation

$$\begin{bmatrix} X \\ 1 \end{bmatrix} = \begin{bmatrix} hA & C \\ 0 & 1 \end{bmatrix} \begin{bmatrix} X_0 \\ 1 \end{bmatrix},$$

where $A \in SO_1(3)$, $A^t = \varepsilon A^{-1} \varepsilon$ and the signature matrix $\varepsilon$ is the diagonal matrix $(\delta_{ij}\varepsilon_j)$ whose diagonal entries are $\varepsilon_1 = -1$, $\varepsilon_2 = \varepsilon_3 = 1$. Hence $\varepsilon^{-1} = \varepsilon = \varepsilon^t$. $\mathbf{X}$ and $\mathbf{X}_0$ are real matrices of $3 \times 1$ type and $h$ is a homothetic scale. $A$, $h$ and $C$ are differentiable functions of $C^\infty$ class of a parameter $t$. $\mathbf{X}$ and $\mathbf{X}_0$ correspond to the position vectors of the same point with respect to the rectangular coordinate frames of the fixed space $R$ and the moving space $R_0$, respectively. At the initial time $t = t_0$, we assume that coordinate systems in $R$ and $R_0$ are coincident [10].

## 3. Split Quaternions and Spacelike Constant Slope Surfaces Lying in the Timelike Cone

In this section, we study unit timelike quaternions with the spacelike vector parts and spacelike constant slope surfaces lying in the timelike cone. Firstly, we obtain the following equation for later use.

A unit timelike quaternion with the spacelike vector part $Q(u,v) = \cosh(\xi_1(u)/2) - \sinh(\xi_1(u)/2) f'(v)$ defines a 2-dimensional surface on $\mathbf{H}_1^3$, where $\xi_1(u) = \coth\theta \ln u$, $\theta$ is a positive constant angle function, $f' = (f_1', f_2', f_3')$ and $f$ is a unit speed spacelike curve on $\mathbf{H}^2$. Thus, using the transformation law $(Q \times V_r \times Q^{-1})_i = \sum_{j=1}^{3} R_{ij}(V_r)_j$, the corresponding rotation matrix can be found as

$$R_Q = \begin{bmatrix} \cosh^2\frac{\xi_1}{2} + \sinh^2\frac{\xi_1}{2}(f_1'^2 + f_2'^2 + f_3'^2) & -2\sinh^2\frac{\xi_1}{2} f_1'f_2' - \sinh\xi_1 f_3' & -2\sinh^2\frac{\xi_1}{2} f_1'f_3' + \sinh\xi_1 f_2' \\ 2\sinh^2\frac{\xi_1}{2} f_1'f_2' - \sinh\xi_1 f_3' & \cosh^2\frac{\xi_1}{2} + \sinh^2\frac{\xi_1}{2}(-f_1'^2 - f_2'^2 + f_3'^2) & -2\sinh^2\frac{\xi_1}{2} f_2'f_3' + \sinh\xi_1 f_1' \\ 2\sinh^2\frac{\xi_1}{2} f_1'f_3' + \sinh\xi_1 f_2' & -2\sinh^2\frac{\xi_1}{2} f_2'f_3' - \sinh\xi_1 f_1' & \cosh^2\frac{\xi_1}{2} + \sinh^2\frac{\xi_1}{2}(-f_1'^2 + f_2'^2 - f_3'^2) \end{bmatrix},$$

(5)

where $r = (S_r, V_r)$.

We now give the relation between unit timelike quaternions with the spacelike vector parts and spacelike constant slope surfaces lying in the timelike cone.

**Theorem 5.** Let $x : M \to \mathbf{E}_1^3$ be a spacelike surface immersed in Minkowski 3-space $\mathbf{E}_1^3$ and $x$ lies in the timelike cone. Then the constant slope surface $M$ can be reparametrized by

$$x(u,v) = Q_1(u,v) \times Q_2(u,v),$$

where $Q_1(u,v) = \cosh\xi_1(u) - \sinh\xi_1(u) f'(v)$ is a unit timelike quaternion with the spacelike vector part, $Q_2(u,v) = u \sinh\theta f(v)$ is a surface and a pure split quaternion in $\mathbf{E}_1^3$ and $f$ is a unit speed spacelike curve on $\mathbf{H}^2$.

*Proof.* Since $Q_1(u,v) = \cosh\xi_1(u) - \sinh\xi_1(u) f'(v)$ and $Q_2(u,v) = u\sinh\theta f(v)$, we obtain

$$\begin{aligned} Q_1(u,v) \times Q_2(u,v) &= \left(\cosh\xi_1(u) - \sinh\xi_1(u) f'(v)\right) \times \left(u\sinh\theta f(v)\right) \\ &= u\sinh\theta \left(\cosh\xi_1(u) - \sinh\xi_1(u) f'(v)\right) \times f(v) \\ &= u\sinh\theta \cosh\xi_1(u) f(v) - u\sinh\theta \sinh\xi_1(u) f'(v) \times f(v). \end{aligned} \quad (6)$$

Using Eq. (3), we have

$$f' \times f = \langle f', f \rangle + f' \wedge f.$$

Since $f$ is a unit speed spacelike curve on $\mathbf{H}^2$, we have

$$-f' \times f = f \wedge f' \quad (7)$$

If we substitute Eq. (7) into Eq. (6), we get

$$Q_1(u,v) \times Q_2(u,v) = u\sinh\theta \left(\cosh\xi_1(u) f(v) + \sinh\xi_1(u) f(v) \wedge f'(v)\right).$$

Therefore applying Eq. (1), we conclude

$$Q_1(u,v) \times Q_2(u,v) = x(u,v)$$

This completes the proof.

From Theorem 5, we can see that the spacelike constant slope surface $M$ lying in the timelike cone is the split quaternion product of 2-dimensional surfaces $Q_1(u,v)$ on $\mathbf{H}_1^3$ and $Q_2(u,v)$ in $\mathbf{E}_1^3$.

Let us consider the relations among rotation matrices $R_Q$, homothetic motions and spacelike constant slope surfaces lying in the timelike cone. We have the following results of Theorem 5.

**Corollary 6.** Let $R_Q$ be the rotation matrix corresponding to the unit timelike quaternion with the spacelike vector part $Q(u,v)$. Then the spacelike constant slope surface $M$ can be written as

$$x(u,v) = R_Q Q_2(u,v).$$

**Corollary 7.** For the homothetic motion $\tilde{Q}(u,v) = u\sinh\theta Q_1(u,v)$, the constant slope surface $M$ can be reparametrized by $x(u,v) = \tilde{Q}(u,v) \times f(v)$. Therefore, we have

$$x(u,v) = u\sinh\theta R_Q f(v) \quad (8)$$

We can give the following remarks regarding Theorem 5 and Corollary 7.

**Remark 8.** Theorem 5 says that both the points and the position vectors on the surface $Q_2(u,v)$ are rotated by $Q_1(u,v)$ through the hyperbolic angle $\xi_1(u)$ about the spacelike axis $Sp\{f'(v)\}$.

**Remark 9.** Corollary 7 says that the position vector of the curve $f(v)$ is rotated by $\tilde{Q}(u,v)$ through the hyperbolic angle $\xi_1(u)$ about the spacelike axis $Sp\{f'(v)\}$ and extended through the homothetic scale $u\sinh\theta$.

## 4. Split Quaternions and Spacelike Constant Slope Surfaces Lying in the Spacelike Cone

In this section, as in Section 3, we study unit timelike quaternions with the spacelike vector parts and spacelike constant slope surfaces lying in the spacelike cone. Similarly, we obtain the following equation for later use.

A unit timelike quaternion with the spacelike vector part $Q(u,v) = \cosh(\xi_2(u)/2) - \sinh(\xi_2(u)/2)g'(v)$ defines a 2-dimensional surface on $\mathbf{H}_1^3$, where $\xi_2(u) = \tanh\theta \ln u$, $\theta$ is a positive constant angle function, $g' = (g_1', g_2', g_3')$ and $g$ is a unit speed spacelike curve on $\mathbf{S}_1^2$. Thus, using the transformation law $(Q \times V_r \times Q^{-1})_i = \sum_{j=1}^{3} R_{ij}(V_r)_j$, the corresponding rotation matrix can be found as

$$R_Q = \begin{bmatrix} \cosh^2\frac{\xi_2}{2} + \sinh^2\frac{\xi_2}{2}(g_1'^2 + g_2'^2 + g_3'^2) & -2\sinh^2\frac{\xi_2}{2}g_1'g_2' - \sinh\xi_2 g_3' & -2\sinh^2\frac{\xi_2}{2}g_1'g_3' + \sinh\xi_2 g_2' \\ 2\sinh^2\frac{\xi_2}{2}g_1'g_2' - \sinh\xi_2 g_3' & \cosh^2\frac{\xi_2}{2} + \sinh^2\frac{\xi_2}{2}(-g_1'^2 - g_2'^2 + g_3'^2) & -2\sinh^2\frac{\xi_2}{2}g_2'g_3' + \sinh\xi_2 g_1' \\ 2\sinh^2\frac{\xi_2}{2}g_1'g_3' + \sinh\xi_2 g_2' & -2\sinh^2\frac{\xi_2}{2}g_2'g_3' - \sinh\xi_2 g_1' & \cosh^2\frac{\xi_2}{2} + \sinh^2\frac{\xi_2}{2}(-g_1'^2 + g_2'^2 - g_3'^2) \end{bmatrix},$$

(9)

where $r = (S_r, V_r)$.

Now we give the relation between unit timelike quaternions with the spacelike vector parts and the spacelike constant slope surfaces lying in the spacelike cone.

**Theorem 10.** Let $x: M \to \mathbf{E}_1^3$ be a spacelike surface immersed in Minkowski 3-space $\mathbf{E}_1^3$ and $x$ lies in the spacelike cone. Then the constant slope surface $M$ can be reparametrized by

$$x(u,v) = Q_1(u,v) \times Q_2(u,v),$$

where $Q_1(u,v) = \cosh\xi_2(u) - \sinh\xi_2(u)g'(v)$ is a unit timelike quaternion with spacelike vector part, $Q_2(u,v) = u\cosh\theta g(v)$ is a surface and a pure split quaternion in $\mathbf{E}_1^3$ and $g$ is a unit speed spacelike curve on $\mathbf{S}_1^2$.

*Proof.* Since $Q_1(u,v) = \cosh\xi_2(u) - \sinh\xi_2(u)g'(v)$ and $Q_2(u,v) = u\cosh\theta g(v)$, we obtain

$$Q_1(u,v) \times Q_2(u,v) = \left(\cosh \xi_2(u) - \sinh \xi_2(u) g'(v)\right) \times \left(u \cosh \theta g(v)\right)$$
$$= u \cosh \theta \left(\cosh \xi_2(u) - \sinh \xi_2(u) g'(v)\right) \times g(v) \qquad (10)$$
$$= u \cosh \theta \cosh \xi_2(u) g(v) - u \cosh \theta \sinh \xi_2(u) g'(v) \times g(v).$$

By using Eq. (3), we have
$$g' \times g = \langle g', g \rangle + g' \wedge g.$$

Since $g$ is a unit speed spacelike curve on $\mathbf{S}_1^2$, we have
$$-g' \times g = g \wedge g' \qquad (11)$$

Substituting Eq. (11) into Eq. (10) gives
$$Q_1(u,v) \times Q_2(u,v) = u \cosh \theta \left(\cosh \xi_2(u) g(v) + \sinh \xi_2(u) g(v) \wedge g'(v)\right).$$

Therefore applying Eq. (2) we conclude
$$Q_1(u,v) \times Q_2(u,v) = x(u,v)$$

This completes the proof.

Similarly, from Theorem 10, we can see that the spacelike constant slope surface $M$ lying in the spacelike cone is the split quaternion product of 2-dimensional surfaces $Q_1(u,v)$ on $\mathbf{H}_1^3$ and $Q_2(u,v)$ in $\mathbf{E}_1^3$.

Let us consider the relations among rotation matrices $R_Q$, homothetic motions and spacelike constant slope surfaces lying in the spacelike cone. We have the following results of Theorem 10.

**Corollary 11.** Let $R_Q$ be the rotation matrix corresponding to the unit timelike quaternion with spacelike vector part $Q(u,v)$. Then the spacelike constant slope surface $M$ can be written as
$$x(u,v) = R_Q Q_2(u,v).$$

**Corollary 12.** For the homothetic motion $\tilde{Q}(u,v) = u \cosh \theta Q_1(u,v)$, the constant slope surface $M$ can be reparametrized by $x(u,v) = \tilde{Q}(u,v) \times g(v)$. Therefore, we have
$$x(u,v) = u \cosh \theta R_Q g(v) \qquad (12)$$

We can give the following remarks regarding Theorem 10 and Corollary 12.

**Remark 13.** Theorem 10 says that both the points and the position vectors on the surface $Q_2(u,v)$ are rotated by $Q_1(u,v)$ through the hyperbolic angle $\xi_2(u)$ about the spacelike axis $Sp\{g'(v)\}$.

**Remark 14.** Corollary 12 says that the position vector of the curve $g(v)$ is rotated by $\tilde{Q}(u,v)$ through the hyperbolic angle $\xi_2(u)$ about the spacelike axis $Sp\{g'(v)\}$ and extended through the homothetic scale $u\cosh\theta$.

## 5. Examples and Their Pictures

We now illustrate some examples of spacelike constant slope surfaces and draw their picture by using Mathematica computer program.

**Example 15.** Let us consider a unit spacelike curve on $\mathbf{H}^2$ defined by
$$f(v) = (\cosh v, 0, \sinh v).$$
Taking $\theta = 7$ and $\coth 7 \cong 1$, the homothetic motion is equal to
$$\tilde{Q}(u,v) = u\sinh 7 \left(\cosh(\ln u) + \left(-\sinh(\ln u)\sinh v, 0, -\sinh(\ln u)\cosh v\right)\right).$$
By using Eq. (8), we have the following spacelike constant slope surface:

$$x(u,v) = u\sinh 7 \begin{bmatrix} \cosh^2\left(\frac{\ln u}{2}\right) + \sinh^2\left(\frac{\ln u}{2}\right)\cosh 2v & -\sinh(\ln u)\cosh v & -\sinh^2\left(\frac{\ln u}{2}\right)\sinh 2v \\ -\sinh(\ln u)\cosh v & \cosh(\ln u) & \sinh(\ln u)\sinh v \\ \sinh^2\left(\frac{\ln u}{2}\right)\sinh 2v & -\sinh(\ln u)\sinh v & \cosh^2\left(\frac{\ln u}{2}\right) - \sin^2\left(\frac{\ln u}{2}\right)\cosh 2v \end{bmatrix} \begin{bmatrix} \cosh v \\ 0 \\ \sinh v \end{bmatrix}$$

and then
$$x(u,v) = \begin{bmatrix} u\sinh 7 \cosh(\ln u)\cosh v \\ -u\sinh 7 \sinh(\ln u) \\ u\sinh 7 \cosh(\ln u)\sinh v \end{bmatrix}.$$

Hence, we can draw the picture of $x(u,v) = u\sinh\theta R_Q f(v)$ as follows:

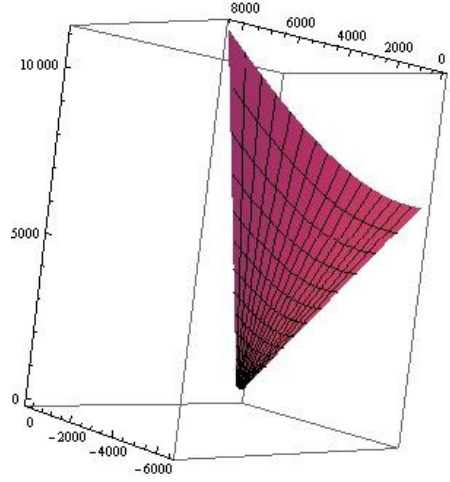

**Figure 1.** Spacelike constant slope surface $x(u,v) = u \sinh\theta R_Q f(v)$,

$$f(v) = (\cosh v, 0, \sinh v), \quad \theta = 7.$$

**Example 16.** Let us consider a unit spacelike curve on $\mathbf{S}_1^2$ defined by

$$g(v) = (0, \cos v, \sin v).$$

Taking $\theta = 7$ and $\tanh 7 \cong 1,$ the homothetic motion is equal to

$$\tilde{Q}(u,v) = u\cosh 7\left(\cosh(\ln u) + (0, \sinh(\ln u)\sin v, -\sinh(\ln u)\cos v)\right).$$

By using Eq. (12), we have the following spacelike constant slope surface:

$$x(u,v) = u\cosh 7 \begin{bmatrix} \cosh(\ln u) & -\sinh(\ln u)\cos v & -\sinh(\ln u)\sin v \\ -\sinh(\ln u)\cos v & \cosh^2\left(\frac{\ln u}{2}\right) + \sinh^2\left(\frac{\ln u}{2}\right)\cos 2v & \sinh^2\left(\frac{\ln u}{2}\right)\sin 2v \\ -\sinh(\ln u)\sin v & \sinh^2\left(\frac{\ln u}{2}\right)\sin 2v & \cosh^2\left(\frac{\ln u}{2}\right) - \sin^2\left(\frac{\ln u}{2}\right)\cos 2v \end{bmatrix} \begin{bmatrix} 0 \\ \cos v \\ \sin v \end{bmatrix}$$

and then

$$x(u,v) = \begin{bmatrix} -u\cosh 7\sinh(\ln u) \\ u\cosh 7\cosh(\ln u)\cos v \\ u\cosh 7\cosh(\ln u)\sin v \end{bmatrix}.$$

Hence, we can draw the picture of $x(u,v) = u\cosh\theta R_Q g(v)$ as follows:

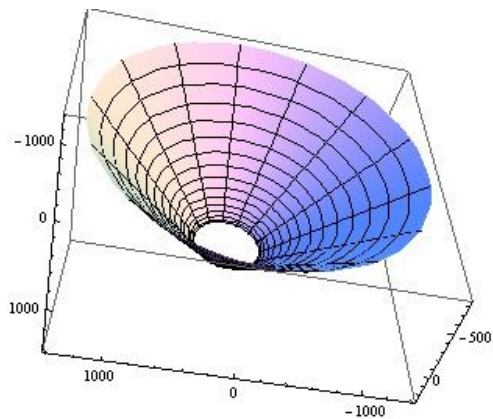

**Figure 2.** Spacelike constant slope surface $x(u,v) = u\cosh\theta R_Q g(v)$,

$$g(v) = (0, \cos v, \sin v), \quad \theta = 7.$$

**References**

[1] M. Babaarslan, Y. Yayli, A new approach to constant slope surfaces with quaternions, ISRN Geometry, In Press.

[2] F. Dillen, J. Fastenakels, J. Van der Veken, L. Vrancken, Constant angle surfaces in $\mathbf{S}^2 \times \mathbf{R}$, Monaths. Math., **152** (2007) 89-96.

[3] F. Dillen, M.I. Munteanu, Constant angle surfaces in $\mathbf{H}^2 \times \mathbf{R}$, Bull. Braz. Math. Soc., **40** (2009) 85-97.

[4] Y. Fu, D. Yang, On constant slope spacelike surfaces in 3-dimensional Minkowski space, J. Math. Anal. Appl. 385 (1) (2012) 208-220.

[5] J. Inoguchi, Timelike surfaces of constant mean curvature in Minkowski 3-space, Tokyo J. Math. **21** (1) 1998 140-152.

[6] L. Kula, Y. Yayli, Split quaternions and rotations in semi Euclidean space $\mathbf{E}_2^4$, J. Korean Math. Soc. **44** (6) (2007) 1313-1327.

[7] R. Lopez, M.I. Munteanu, Constant angle surfaces in Minkowski space, Bull. Belg. Math. Soc. - Simon Stevin, **18** (2) (2011) 271-286.

[8] B. O'Neill, Semi-Riemannian Geometry, Pure and Applied Mathematics, 103, Academic Press, Inc. [Harcourt Brace Jovanovich, Publishers], New York, 1983.

[9] M. Ozdemir, A.A. Ergin, Rotations with unit timelike quaternions in Minkowski 3-space, J. Geom. Phys. **56** (2) (2006) 322-326.

[10] M. Tosun, A. Kucuk, M.A. Gungor, The homothetic motions in the Lorentzian 3-space, Acta Math. Sci. **26B** (4) (2006) 711-719.

Murat Babaarslan
Department of Mathematics
Bozok University
66100, Yozgat
Turkey
e-mail: murat.babaarslan@bozok.edu.tr

Yusuf Yayli
Department of Mathematics
Ankara University
06100, Ankara
Turkey
e-mail: yayli@science.ankara.edu.tr